# Silver Hardening via Hypersonic Impacts

*Eliezer Fernando Oliveira[1,2], Pedro Alves da Silva Autreto[1,3] and Douglas Soares Galvão[1,2]*

[1] Applied Physics Department and Center of Computational Engineering and Science, University of Campinas - UNICAMP, Campinas-SP 13083-959, Brazil.

[2] Center for Computational Engineering & Sciences (CCES), University of Campinas - UNICAMP, Campinas, SP, Brazil.

[3] Federal University of ABC, Center of Natural Human Science, Santo Andre-SP 09210-580, Brazil.

ABSTRACT

*The search for new ultra strong materials has been a very active research area. With relation to metals, a successful way to improve their strength is by the creation of a gradient of nanograins (GNG) inside the material. Recently, R. Thevamaran et al. [Science v354, 312-316 (2016)] propose a single step method based on high velocity impact of silver nanocubes to produce high-quality GNG. This method consists of producing high impact collisions of silver cubes at hypersonic velocity (~400 m/s) against a rigid wall. Although they observed an improvement in the mechanical properties of the silver after the impact, the GNG creation and the strengthening mechanism at nanoscale remain unclear. In order to gain further insights about these mechanisms, we carried out fully atomistic molecular dynamics simulations (MD) to investigate the atomic conformations/rearrangements during and after high impact collisions of silver nanocubes at ultrasonic velocity. Our results indicate the co-existence of polycrystalline arrangements after the impact formed by core HCP domains surrounded by FCC ones, which could also contribute to explain the structural hardening.*

**INTRODUCTION**

Ultra-strong materials have been used in several technological applications, such as protective materials industry and armors for vehicles, aircraft, etc [1]. In the case of metals, one of the most successful methods to create ultra-strong metallic structure is through gradient creation of nanograins (GNG) [2,3]. However, in general GNC require a multi-step process and can result in large number of grains with micro instead of nanosize domains [4]. Recently, R. Thevamaran *et al.* [5] developed a new method in which using a single-step process it was possible to produce almost perfect nanoscale silver GNG. This method consists of shooting silver microcubes at hypersonic velocity (~400 m/s) against solid targets. This process is very efficient to enhance their mechanical (especially hardening) properties. But, the stored elastic energy in the deformed silver nanostructures promotes a recrystallization and they undergo to a transition from nano-polycrystalline to monocrystalline one, leading to structures quite similar to the ones before impacts [5]. About 40 days after the impacts, no GNG is observed in the structure, which compromises the induced hardening and other mechanical properties. An atomistic investigation of the impact dynamics and structural deformation during/after impact can provide further insights on these mechanisms and provide useful information in order to prevent the recrystallization to occur. In this work, we carried out fully atomistic molecular dynamics simulations of silver nanocubes shot at hypersonic velocity against a solid target.

**MATERIAL AND METHODS**

We carried out molecular dynamics (MD) simulations using the Embedded-Atom Method potential (EAM) [6,7] as implemented in the computational code Large-scale Atomic/Molecular Massively Parallel Simulator (LAMMPS) [8]. We have considered face centered cubic (FCC) Ag cubes of sizes of $20a_0$, where $a_0$ is the lattice parameter equals to 3.52 Å [9]. These cubes contain ~32000 atoms. The geometry of the silver cube was optimized using minimum energy conjugate gradient technique, followed by a thermal equilibration at 10K in a NVT ensemble. After this, NVE ensemble was considered and the cubes were shot against an impenetrable target composed of a Lennard-Jones wall. We used a shooting velocity for the silver cube of 400 m/s, the same employed by R. Thevamaran *et al.* [5]. We have considered three different cube-shooting orientations towards to targets: [100] (face), [110] (lateral) and [111] (corner) of silver cubes. For all MD runs, a time step of 0.025 fs and the Nosé-Hoover thermostat were used. This methodology has been successfully used in studies of nanostructures under high velocity impacts [10-12]. The stress values were calculated by computing the forces on each atom, then obtaining the stress tensor and from this we calculated the von Mises stress [13], which is the second invariant of the deviatoric stress tensor. By calculating the von Mises stress for each atom at every 100 time-steps, it was possible to visualize the time evolution of the local and average stresses, which allowed us to obtain a detailed information on the deformation processes.

**RESULTS AND DISCUSSIONS**

We present the average accumulated stress of these simulations since the impact moment in Figure 1(a). Some representative MD snapshots from face, lateral and corner impact orientations are also shown in Fig. 1(b). Stresses at the lateral and corner impacts reach higher values than those from face impacts, which means that is more difficult to deform the silver cube at the lateral and corner than at face impacts. While the face impact efficiently absorbs the kinetic energy through structural deformations (see Figure 1(b)), for the lateral and corner cases this is obtained through mass loss/atom ejection, and the smashing process will finish firstly for the face direction (see Figure 1(b)).

As discussed before, it was experimentally observed that the smashed silver structures recrystallize after a number of days [5]. In order to mimic this aging-like process, we perform an energy minimization of all post-impacted structures. As we obtain similar results for all cube shooting directions, we will only presents the results for the face one. A re-optimized structure for the face impact case is presented in Figure 2(a). A visual inspection of the annealed post-impacted structures shows partial crystallinity recovery after the minimization process. A cross-section view (Figure 2(b)) indicates that the recrystallization occurs only in the top border and at center of the deformed structure, while the bottom remained amorphous-like. The important differences between the crystallinity of bottom and center regions is that the top borders recover up to FCC structure (orange in Figure 2(b)) and the center presents hexagonal close-packing (HCP) (red in Figure 2(b)). These observations are corroborated by the radial pair distribution function (g(r)) analyses for three regions of the post-minimized structure: top border, center and bottom; these results are presented in Figure 2(c) As seen, the bottom region (dashed line) remains amorphous, indicating that even a long-time relaxation after the impact is not enough to recover any crystalline phase. For center (dotted line) and top border (full line) regions it can be noted only in the fourth peak, that exhibits two distinct crystalline phases: HCP for center and FCC for top border regions [14].

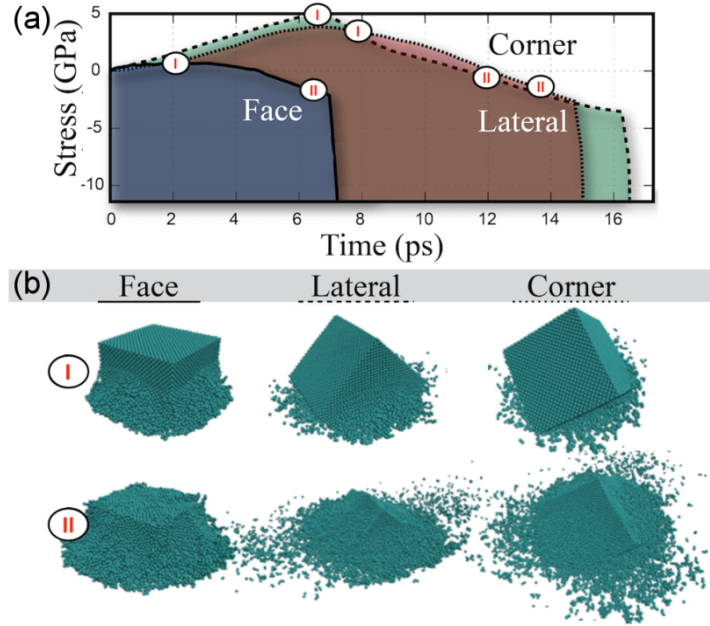

Figure 1: (a) Average accumulated stress during the impact at face, lateral and corner, respectively. (b) Representative MD snapshots of different stages of impacts. I and II correspond to impact stages of high and partially released stress values, respectively.

The FCC to HCP phase transition, as well as their coexistence are well-known in the literature and the most stable phase for Ag is the FCC, while HCP is a metastable one [15]. The coexistence of FCC and HCP domains could contribute to the observed Ag hardening and this possibility was not considered by R. Thevamaran *et al.* [5]. Our results suggest that the effect of high impact velocity could be beyond the grain size modifications, once it can provide a phase transition inside each grain. These modifications might affect the silver mechanical properties. In order to test this hypothesis we performed a strain-stress analysis for silver nanostructures in a pure FCC and a mix FCC+HCP. We use structures with same size ($18a_0 \times 20a_0 \times 12a_0$) and the mix FCC+HCP Ag structure were obtained 'extracting' core samples from post-impact minimized face smashed Ag cubes. After a pre-optimization of the structure, the strain-stress curve was obtained by applying a compressive loading force along the z direction;

the result is presented in Figure 3. As can be seen, our results suggests that FCC+HCP mixed structure is more resilient to structural deformations than a pure FCC one, since the estimated ultimate strength is about 5.6 GPa for HCP+FCC mixture, in comparison to just 2.5 GPa for the pure FCC.

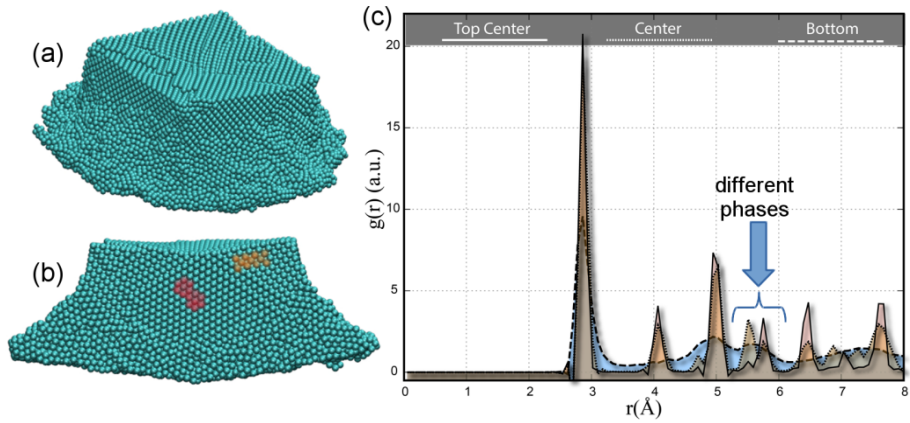

Figure 2: (a) Post-impact minimized structures and its cross-section view (b). The Figure inset highlight the coexistence of FCC and HCP domains; (c) pair radial distribution g(r) for top border, center and a bottom regions, respectively.

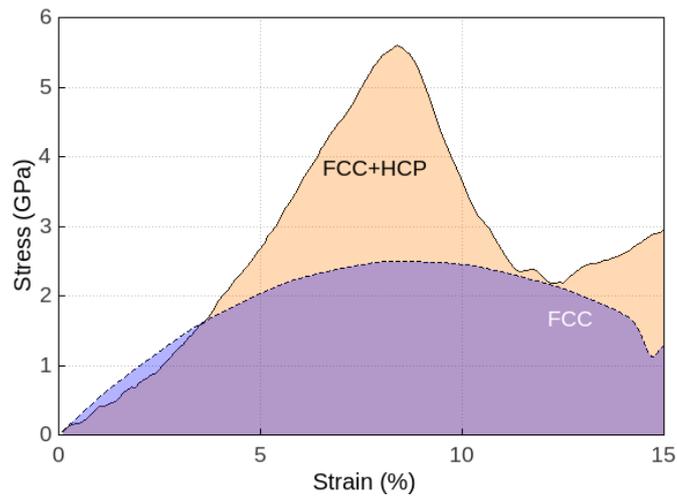

Figure 3: Stress x strain curves for Ag FCC and Ag FCC+HCP (HCP cores surrounded by FCC).

Since silver in HCP arrangement is a metastable structure, it can be spontaneously convert back to FCC because in time the stored energy after the impact could be enough to overcome the interconversion (HCP to FCC) energy barrier. This effect combined with GNG coalescence cause a monocrystalline formation after a long-time relaxation and compromise the good mechanical properties observed in the smashed Ag nanocubes, which is in good agreement with the experimental observations [5].

**CONCLUSIONS:**

Our MD results indicate the co-existence of HCP/FCC domains after silver cubes impact. The association of HCP cores surrounded by FCC structures into silver improves its mechanical properties, in which higher ultimate strengths are obtained in comparison to the pure silver FCC. This suggests that besides the proposed gradient nanograin formation to explain the increased hardness, these HCP/FCC domains can also contribute to silver hardening. As the HCP phase in silver is a metastable state and eventually can return to the FCC phase, a way to produce a strong material could be to employ a material or alloy containing at least two stable crystalline phases. This can be a new way to design ultra-strong materials, even in the absence of GNC domains [16].

**ACKNOWLEDGMENTS:**

We would like to thank the Brazilian agency FAPESP (Grants 2013/08293-7 and 2016/18499-0) for financial support and Center for Computational Engineering and Sciences (CCES) for the computational support.

**REFERENCES**

1. T. W. McAllister, J. C. Ford, S. Ji, J. G. Beckwith, L. A. Flashman, K. Paulsen and R. M. Greenwald, *Ann. Biomed. Eng.* **40**, 127 (2012).
2. M. D. Uchic, D. M. Dimiduk, J. N. Florando and W. D. Nix, *Science* **305**, 986 (2004).
3. J. R. Greer and J. T. M. De Hosson, *Prog. Mater. Sci.* **56**, 654–724 (2011).
4. K. Lu, *Science* **345**, 1455 (2014).
5. R. Thevamaran, O. Lawal, S. Yazdi, S.-J. Jeon, J.-H. Lee and E. L. Thomas, *Science* **354**, 312 (2016).
6. M. S. Daw and M. I. Baskes, *Phys. Rev. B* **29**, 6443 (1984).
7. M. S. Daw, S. M. Foiles and M. I. Baskes, *Mater. Sci. Rep.* **9**, 251 (1993).
8. S. J. Plimpton, *J. Comput. Phys.* **117**, 1 (1995).
9. J. F. Shackelford, *Introduction to Materials Science for Engineers*, Pearson, London, 2015.
10. S. Ozden, P. A. S. Autreto, C. S. Tiwary, S. Khatiwada, L. Machado, D. S. Galvao, R. Vajtai, E. V. Barrera and P. M. Ajayan, *Nano Lett* **14**, 4131 (2014).
11. L. D. Machado, S. Ozden, C. S. Tiwary, P. A. S. Autreto, R. Vajtai, E. V. Barrera, D. S. Galvao and P. M. Ajayan, *Phys. Chem. Chem. Phys.* **18**, 14776 (2016).
12. S. Ozden, L. D. Machado, C. S. Tiwary, P. A. S. Autreto, R. Vajtai, E. V. Barrera, D. S. Galvao and P. M. Ajayan, *ACS Appl. Mater. Interfaces* **8**, 24819 (2016).
13. A. Zang and O. Stephansson, *Stress Field of the Earth's Crust*, Springer, Houten, 2009.
14. Z. Y. Hou, K. J. Dong, Z. A. Tian, R. S. Liu, Z. Wang and J. G. Wang, *Phys. Chem. Chem. Phys.* **18**, 17461 (2016).
15. N. Zhou, D. Li and D. Yang, *Nanoscale Res. Lett.* **9**, 302 (2014).
16. E. F. Oliveira, P. A. S. Autreto and D. S. Galvao, *to be published.*